# Multimodal assessment of nigrosomal degeneration in Parkinson's disease

Jason Langley [a], Daniel E. Huddleston [b], Bruce Crosson [b,c,d], David Song [e], Stewart A. Factor [b], and Xiaoping Hu [a,f]*

a. Center for Advanced Neuroimaging, University of California, Riverside, Riverside, CA, USA
b. Department of Neurology, Emory University, Atlanta, GA, USA
c. Department of Veterans Affairs Center for Visual and Neurocognitive Rehabilitation, Atlanta Veterans Affairs Medical Center, Decatur, GA, USA
d. Department of Psychology, Georgia State University, Atlanta, GA, USA
e. Department of Neurology, University of California, Riverside, Riverside, CA, USA
f. Department of Bioengineering, University of California, Riverside, Riverside, CA, USA

*Correspondence to:
Xiaoping P. Hu, Ph.D.
Professor and Chair Department of Bioengineering
University of California, Riverside
xhu 'at' engr.ucr.edu

## Abstract

**Background**: Approximately forty percent of all dopaminergic neurons in SNpc are located in five dense neuronal clusters, named nigrosomes. $T_2$- or $T_2^*$-weighted images are used to delineate the largest nigrosome, named nigrosome-1. In these images, nigrosome-1 is a hyperintense region in the caudal and dorsal portion of the $T_2$- or $T_2^*$-weighted substantia nigra. In PD, nigrosome-1 experiences iron accumulation, which leads to a reduction in $T_2$-weighted hyperintensity. Here, we examine neuromelanin-depletion and iron deposition in regions of interest (ROIs) derived from quantitative-voxel based morphometry (qVBM) on neuromelanin-sensitive images and compare the ROIs with nigrosome-1 identified in $T_2^*$-weighted images.

**Methods:** Neuromelanin-sensitive and multi-echo gradient echo imaging data were obtained. $R_2^*$ was calculated from multi-echo gradient echo imaging data. qVBM analysis was performed on neuromelanin-sensitive images and restricted to SNpc. Mean neuromelanin-sensitive contrast and $R_2^*$ was measured from the resulting qVBM clusters. Nigrosome-1 was segmented in $T_2^*$-weighted images of control subjects and its location was compared to the spatial location of the qVBM clusters.

**Results:** Two bilateral clusters emerged from the qVBM analysis. These clusters showed reduced neuromelanin-sensitive contrast and increased mean $R_2^*$ in PD as compared to controls. Cluster-1 from the qVBM analysis was in a similar spatial location as nigrosome-1, as seen in $T_2^*$-weighted images.

**Conclusion:** qVBM cluster-1 shows reduced neuromelanin-sensitive contrast and is in a similar spatial position as nigrosome-1. This region likely corresponds to nigrosome-1 while the second cluster may correspond to nigrosome-2.

## Introduction

The substantia nigra is a paired midbrain structure comprised of two substructures, substantia nigra pars compacta (SNpc) and substantia nigra pars reticulata. SNpc contains dense clusters of neuromelanin-containing dopaminergic neurons [1]. SNpc is profoundly affected in the prodromal stages of Parkinson's disease (PD) where up to 50% of melanized neurons in SNpc are lost by the time of clinical presentation [2].

The anatomy of SNpc and the distribution of its dopaminergic neurons can be organized based on calbindin $D_{28K}$ immunostaining. Within SNpc there are five $D_{28K}$ negative subregions containing clusters of melanized dopamine neurons, labeled nigrosomes 1-5. Those subregions that are calbindin $D_{28K}$ positive lack these dense clusters of melanized neurons and are referred to as matrix [3]. The largest nigrosome, nigrosome-1, exhibits the most profound PD-related neurodegeneration with up to 98% neuronal loss. The SNpc region with the second highest degree of neuronal loss is nigrosome-2 in PD, followed by nigrosomes-4, 3, and 5 [4]. Other work found the greatest loss of dopaminergic neurons centered in the lateral and ventral portions of SNpc, a subregion containing nigrosome-1 [5, 6].

Iron deposition has been found in conjunction with PD-related neuronal loss in SNpc [7]. $T_2^*$-weighted images are sensitive to iron and regions with high iron content appear hypointense in these images and these images have been employed to detect loss of hyperintensity associated with iron deposition in nigrosome-1. In $T_2^*$-weighted images, nigrosome-1 can be seen as a relatively hyperintense region in the caudal and dorsal portion of the $T_2^*$-hypointense substantia nigra in control subjects [8]. This results in an overall appearance resembling a swallow tail [8]. However, hypointense signal is observed in this region in PD, and the swallow-tail appearance is lost due to neurodegeneration-associated iron deposition [9-11].

$T_2^*$-weighted images are not sensitive to neuromelanin, and therefore accurate delineation of SNpc is difficult in these images [12]. Incidental or explicit magnetization transfer effects can be used to generate neuromelanin-sensitive contrast and delineate SNpc [13, 14]. Studies localizing SNpc with neuromelanin-sensitive contrast found a reduction in neuromelanin-sensitive signal [15] or increased iron deposition [16] in the lateral-ventral region of SNpc. The results of these studies may reflect neurodegeneration in nigrosome-1. However, the spatial relationship between nigrosome-1 and the SNpc defined with neuromelanin-sensitive contrast, and the manifestations of nigrosomal neurodegeneration in neuromelanin-sensitive images are largely unexplored.

In this work, a voxel-based analysis is used to determine spatial locations in SNpc most sensitive to PD-related neuronal loss. We further examine the location of nigrosome-1 with respect to the neuromelanin-sensitive substantia nigra (SNpc) and use nigrosome-1 to interpret significant clusters from the voxel-based analysis.

## Materials and Methods

Sixty-one subjects (31 PD and 30 control) participated in this study. Data from 3 PD and 2 control subjects were excluded due to motion artifacts resulting in a final sample size of 56 subjects (28 PD and 28 control). All subjects participating in the study gave written informed consent in accordance with local institutional review board regulations. PD subjects were recruited from the Emory University Movement Disorders Clinic and clinically diagnosed with PD according to the UK Brain Bank criteria [17]. PD patients had early to moderate disease with a Unified Parkinson's Disease Rating Scale Part III (UPDRS-III) motor score of ≤25 in the ON medication state. Control subjects were recruited from a cohort of individuals without major neurological diagnoses followed by the Emory Alzheimer's Disease Research Center. Specific exclusion criteria included the following: 1) patients showing symptoms or signs of secondary or atypical parkinsonism [18], 2) controls were excluded if they scored ≤26 on the Montreal Cognitive Assessment (MOCA) indicating cognitive impairment, 3) any history of vascular territorial stroke, epilepsy, multiple sclerosis, neurodegenerative disease (aside from PD), peripheral neuropathy with motor deficits, parenchymal brain tumor, hydrocephalus, or schizophrenia, 4) treatment with an antipsychotic drug (other than quetiapine at a dose less than 200mg daily), or 5) if there were any contraindications to MRI imaging.

Demographic information including gender, age and education, was collected for each subject and is shown in Table 1. Participants in both the PD and control groups underwent UPDRS-III examination by a fellowship-trained movement disorders neurologist. PD patients were examined and underwent imaging in the ON medication state.

| Variable | HC (n=28) | PD (n=28) | $p$ Value |
|---|---|---|---|
| Gender (M/F) | 17/11 | 15/13 | 0.45 |
| Age (yrs) | 64.8±5.2 | 62.3±9.2 | 0.22 |
| Education (yrs) | 16.8±2.6 | 15.9±2.9 | 0.30 |
| L-DOPA equivalent | — | 626±448 | — |
| UPDRS-III score | 2.4±2.3 | 22.4±8.0 | $<10^{-4}$ |
| Hoehn and Yahr | — | 2.0±0.4 | — |
| Disease Duration (yrs) | — | 4.7±4.5 | — |

Table 1. Demographic information and clinical characteristics of PD patients and healthy controls in the cohort used in this analysis. Data are presented as mean ± standard deviation. L-DOPA – levodopa; UPDRS - Unified Parkinson's Disease Rating Scale.

*Image Acquisition*

Data were acquired on a 3T MRI scanner (Prisma Fit, Siemens Healthineers, Malvern, PA) using a 64-channel receive only coil. Images from a $T_1$-weighted MP-RAGE sequence (echo time (TE)/repetition time (TR)/inversion time=3.02/2600/800 ms, flip angle (FA)=8°, voxel size=0.8×0.8×0.8 mm$^3$) were used for registration from subject space to common space. Neuromelanin-sensitive data were acquired using a magnetization-prepared 2D gradient recalled echo sequence: TE/TR=3.10/354 ms, 416×512 imaging matrix, voxel size=0.39×0.39×3 mm$^3$, 15 slices, FA = 40°, 7 measurements, magnetization transfer preparation pulse (300°, 1.2 kHz off-resonance) [13]. The seven measurements were saved individually for offline processing. Multiecho data were collected with a 6-echo 3D gradient recalled echo sequence: $TE_1/\Delta TE/TR$=4.92/4.92/50 ms, FOV=220×220 mm$^2$, matrix size of 448×336×80, slice thickness=1 mm, and GRAPPA acceleration factor=2.

*Standard space transformation*

Imaging data were analyzed with FMRIB Software Library (FSL). A transformation was derived from individual subject space to Montreal Neurological Institute (MNI) $T_1$-weighted common space using FMRIB's Linear Image Registration Tool and FMRIB's Nonlinear Image Registration Tool in the FSL software package [19]. The procedure for this transformation is as follows: first, the $T_1$-weighted image was skull stripped using the brain extraction tool in FSL. Next, brain extracted $T_1$-weighted images were aligned with the MNI brain extracted image using an affine transformation. Finally, a nonlinear transformation was used to generate a transformation from individual $T_1$-weighted images to $T_1$-weighted MNI common space.

*Neuromelanin SNpc atlas*

A SNpc neuromelanin atlas in MNI space was used in this study to localize SNpc for the voxel-based analysis. The atlas was created from a cohort of 76 healthy older participants (mean age: 66.6 years ± 6.4 years) using a process similar to those outlined in [16].

*$R_2^*$ map creation*

$R_2^*$ is defined as the inverse of the transverse relaxation rate ($1/T_2^*$) and is measured from a multi-echo gradient echo pulse sequence. $R_2^*$ varies linearly with iron content and can be used as a marker of brain iron *in vivo* [20, 21]. $R_2^*$ values were estimated voxel-wise using a custom script in MATLAB by fitting a monoexponential model to the gradient echo images using the following equation:

$$S_i(TE) = S_0 \exp(-R_2^* TE) \qquad [1]$$

where $S_0$ denotes a fitting constant and $S_i$ denotes the signal of a voxel at the *i*th echo time. The resulting $R_2^*$ map was aligned to the $T_1$-weighted image using a transform derived from the magnitude image from the first echo. Finally, mean $R_2^*$ was calculated in the clusters from the quantitative voxel based analysis described in later sections.

*Neuromelanin-sensitive image processing*

Imaging data in this section were analyzed with FSL. First, images from the seven measurements with neuromelanin-sensitive contrast from the magnetization-transfer prepared gradient echo sequence were registered to the first image using a linear transformation tool and averaged. The averaged image was used in the subsequent analysis. A transformation between the resulting magnetization transfer prepared gradient echo image and $T_1$-weighted image was derived using a rigid body transform with boundary-based registration cost function. The quality of registration was checked for each subject and no significant misregistrations were observed across the sample.

Contrast from the magnetization transfer preparation pulse, denoted magnetization transfer contrast (MTC), was estimated using the following equation:

$$MTC = (I - I_{ref})/I_{ref}$$

where *I* denotes the intensity of a voxel in the magnetization transfer prepared gradient echo image and $I_{ref}$ is the mean intensity of a reference region in the magnetization transfer prepared gradient echo image. To ensure consistent placement of reference region in the magnetization transfer prepared gradient echo images across subjects, a reference region was drawn in the cerebral peduncle in MNI $T_1$-weighted common space and then transformed to individual magnetization transfer prepared gradient echo images. A typical registration is seen in supplemental Figure S1. The resulting MTC maps were then transformed to MNI common space for voxel-based morphometry analysis.

*Quantitative Voxel Based Analysis*

In contrast to standard voxel-based morphometry analysis, which uses $T_1$-weighted images to ascertain regions exhibiting reduced $T_1$-contrast, quantitative voxel-based morphometry (qVBM) is applied to semi-quantitative MRI images, such as MTC images (i.e. neuromelanin-sensitive images), and allows for spatial regions exhibiting disease-related changes to be ascertained. Voxel-wise differences in MTC, generated and transformed to common space in the previous section, were found using qVBM [22]. To correct for multiple comparisons over space, we used permutation-based non-parametric inference with the framework of the general linear model with 5000 permutations. Results were considered significant for $p<0.05$ (family-wise error corrected) after initial-cluster forming thresholding at *p*-corrected=0.05. For this analysis, statistical testing was restricted to the control neuromelanin-sensitive SNpc volume. Mean MTC and $R_2^*$ were calculated in the resulting clusters.

*Nigrosome-1*

Nigrosome-1 was manually segmented in images acquired during the 5th echo of the multi-echo gradient echo sequence (TE=24 ms) for control subjects. Nigrosome-1 was defined to be the hyperintense region in slices 2 mm below the lower extent of the red nucleus. Nigrosome-1 for a representative control subject is shown in Figure 1A. For each control, nigrosome-1 was transformed into MNI space using the procedures defined above and then averaged to create a population atlas for nigrosome-1. The nigrosome-1 atlas was then thresholded at a level of 0.5, corresponding to 50% of the population in agreement for nigrosome-1, and binarized.

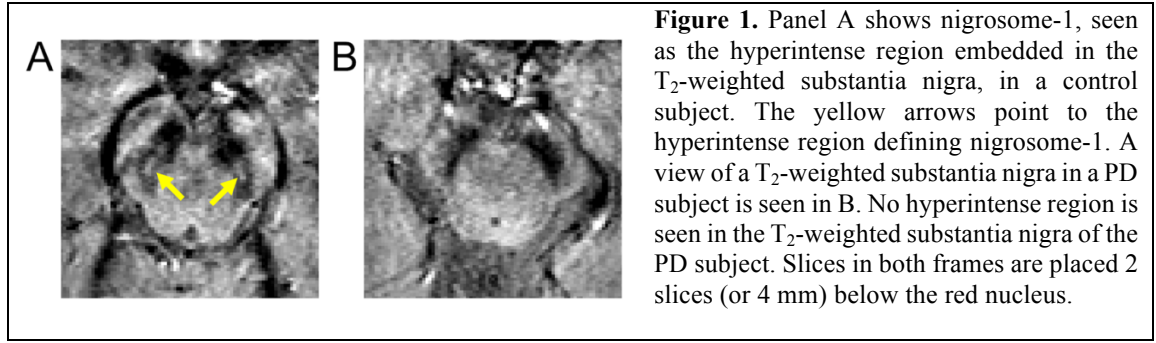

**Figure 1.** Panel A shows nigrosome-1, seen as the hyperintense region embedded in the $T_2$-weighted substantia nigra, in a control subject. The yellow arrows point to the hyperintense region defining nigrosome-1. A view of a $T_2$-weighted substantia nigra in a PD subject is seen in B. No hyperintense region is seen in the $T_2$-weighted substantia nigra of the PD subject. Slices in both frames are placed 2 slices (or 4 mm) below the red nucleus.

The Dice similarity coefficient (DSC) was used to assess the closeness nigrosome-1 and the first cluster from the qVBM analysis. DSC values of 0 and 1 indicate no overlap and perfect overlap between two regions, respectively. DSC is defined as

$$\text{DSC} = \frac{2 * \text{volume}(NS1 \cap C1)}{\text{volume}(NS1) + \text{volume}(C1)}$$

where NS1 denotes nigrosome-1 in MNI space and C1 denotes cluster 1 from the qVBM analysis.

*Statistical Analysis*

All statistical analyses were performed using IBM SPSS Statistics software version 24 (IBM Corporation, Somers, NY, USA) and results are reported as mean ± standard deviation. As nigrosomes exhibit the most profound PD-related neurodegeneration [4], we hypothesize clusters showing significant reductions in MTC from the qVBM analysis spatially overlap with nigrosomes. To correct for multiple comparisons over space in the qVBM analysis, we used permutation-based non-parametric inference with the framework of the general linear model with 5000 permutations [23]. Within each cluster, group SNpc $R_2^*$ and MTC comparisons between PD patients and controls were made using a one-tailed *t*-test. A one-tailed *t*-test is used since the direction of expected effects is known based on prior studies, which reported that PD is associated with MTC reduction and $R_2^*$ increase in SNpc [15, 24]. Linear regressions were performed with mean $R_2^*$ and MTC from qVBM clusters with UPDRS-III score and disease duration. A *p*-value of 0.05 was considered significant for all statistical tests performed in this work. Receiver operator characteristic (ROC) analysis were performed for mean $R_2^*$ and MTC values in each cluster.

**Results**

There was no difference in age (*p*=0.22), educational (*p*=0.30), or sex (*p*=0.25) distribution between PD patients and control subjects. Demographic information is summarized in Table 1.

*qVBM analysis*

Analysis of MTC in SNpc with qVBM revealed two bilateral statistically significant clusters of reduced MTC in PD patients as compared to controls. The clusters are shown in Figure 2B. Cluster 1 was located in the lateral and posterior portions of SNpc (left center of mass: Z=-8.8 mm, Y= -22.2 mm, Z= -15.6 mm; right center of mass: X=-10.2 mm, Y=-22.2 mm, Z=-15.1 mm).

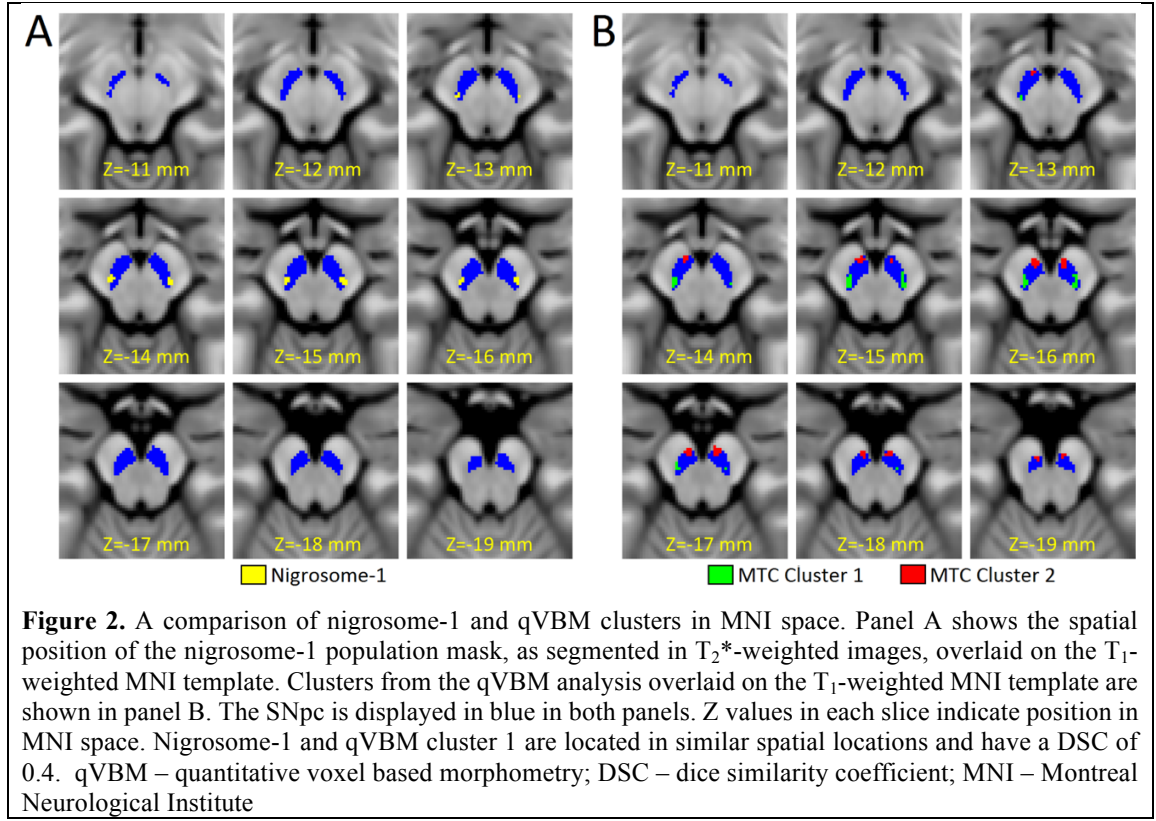

**Figure 2.** A comparison of nigrosome-1 and qVBM clusters in MNI space. Panel A shows the spatial position of the nigrosome-1 population mask, as segmented in $T_2^*$-weighted images, overlaid on the $T_1$-weighted MNI template. Clusters from the qVBM analysis overlaid on the $T_1$-weighted MNI template are shown in panel B. The SNpc is displayed in blue in both panels. Z values in each slice indicate position in MNI space. Nigrosome-1 and qVBM cluster 1 are located in similar spatial locations and have a DSC of 0.4. qVBM – quantitative voxel based morphometry; DSC – dice similarity coefficient; MNI – Montreal Neurological Institute

In cluster 1, mean MTC in the PD group was reduced relative to that of the control group (Control: 0.16±0.03; PD: 0.13±0.03; $p=5.0\times10^{-4}$). Cluster 2 was located in the anterior and medial portions of SNpc (left center of mass: X=-4.6 mm, Y=-16.5 mm, Z=-16.9 mm; right center of mass: x= 6.0 mm, y= -15.7 mm, z= -16.0 mm), and in cluster 2 MTC was also lower in the PD group (Control: 0.18±0.02; PD: 0.15±0.02; $p=8.6\times10^{-5}$). These results are shown in Figure 3. No correlation was observed between ON UPDRS-III score and mean MTC in qVBM cluster 1 ($p=0.412$, $r=0.161$) or between ON UPDRS-III score and mean MTC in qVBM cluster 2 (p=0.065; $r=0.353$).

Iron deposition in the clusters from the qVBM analysis was examined by measuring mean $R_2^*$ within each cluster. Increases in $R_2^*$ were observed in qVBM cluster 1 (Control: 26.0 $s^{-1}$±5.4 $s^{-1}$; PD: 30.2 $s^{-1}$±5.7 $s^{-1}$; $p=0.004$) and qVBM cluster 2 (Control: 33.4 $s^{-1}$±5.6 $s^{-1}$; PD: 37.0 $s^{-1}$±5.7 $s^{-1}$; $p=0.012$) of the PD group as compared to controls. No correlation was observed between ON UPDRS-III score and mean $R_2^*$ in VBM cluster 1 ($p=0.241$, $r=0.127$) or between ON UPDRS-III score and mean $R_2^*$ in VBM cluster 2 ($p=0.109$, $r=0.220$).

In the ROC analysis, mean MTC outperformed mean $R_2^*$ as a marker for PD in qVBM clusters 1 and 2. Area under the curve (AUC) for mean MTC in VBM clusters 1 and 2 was 0.746 (standard error (SE)=0.066; 95% confidence interval (CI)=0.616-0.876; $p=0.002$) and 0.786 (SE=0.062; 95% CI=0.664-0.907; $p=0.0002$), respectively. AUC for mean $R_2^*$ in qVBM cluster 1

was 0.691 (SE=0.066; 95% CI=0.555-0.827; $p$=0.012). Mean $R_2^*$ in qVBM cluster 2 had an AUC of 0.661 (SE=0.073; 95% CI=0.518-0.804; $p$=0.035).

*Nigrosome-1 in $T_2^*$-weighted images*

In $T_2^*$-weighted images, nigrosome-1 is seen as a hyperintense region embedded in the $T_2^*$-weighted substantia nigra [8]. Nigrosome-1 was observed in 27 of the 28 controls participating in this study, but was not visible in 26 of the 28 subjects in the PD group. After transformation of the control nigrosome-1 masks to MNI space, thresholding and binarization, the center of mass for the control nigrosome-1 in MNI space was (X=10.9 mm, Y=-21.6 mm, Z=-14.6 mm) and (X=-10.1 mm, Y=-22.0 mm, Z=-14.9 mm) for the right and left sides (see Figure 2A), respectively. Interestingly, nigrosome-1 and cluster 1 from the qVBM analysis are located in highly similar anatomical positions, with a difference in center of mass between nigrosome-1 and MTC cluster 1 of 1.4 mm and 1.1 mm for the left and right sides, respectively. The DSC for the binarized nigrosome-1 mask and qVBM cluster 1 is 0.40, indicating overlap between qVBM cluster 1 and nigrosome-1.

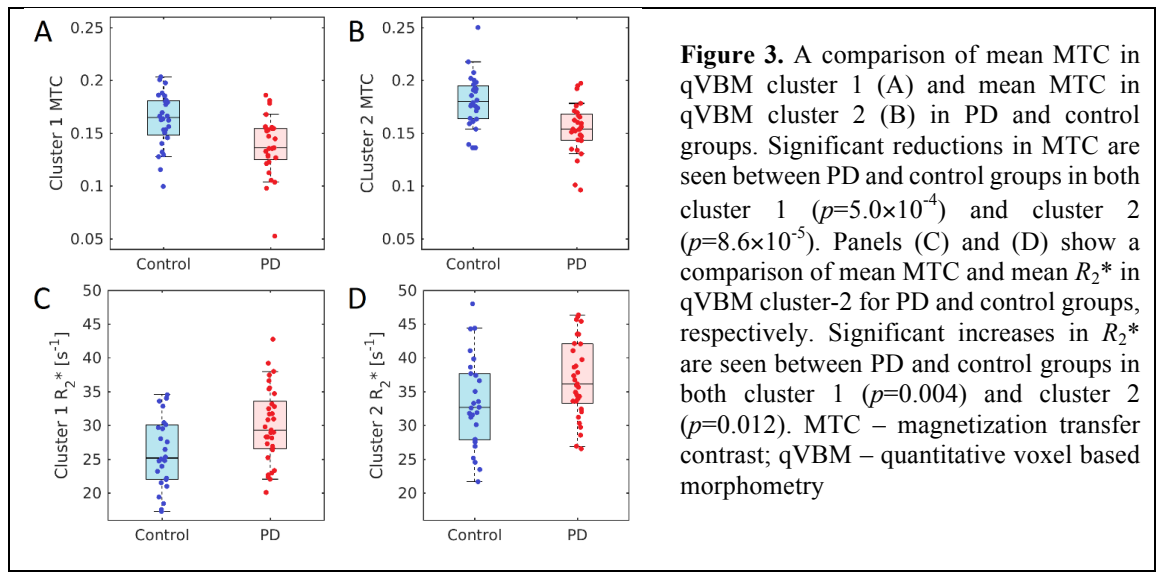

**Figure 3.** A comparison of mean MTC in qVBM cluster 1 (A) and mean MTC in qVBM cluster 2 (B) in PD and control groups. Significant reductions in MTC are seen between PD and control groups in both cluster 1 ($p$=5.0×10$^{-4}$) and cluster 2 ($p$=8.6×10$^{-5}$). Panels (C) and (D) show a comparison of mean MTC and mean $R_2^*$ in qVBM cluster-2 for PD and control groups, respectively. Significant increases in $R_2^*$ are seen between PD and control groups in both cluster 1 ($p$=0.004) and cluster 2 ($p$=0.012). MTC – magnetization transfer contrast; qVBM – quantitative voxel based morphometry

**Discussion**

Delineation of nigrosome-1 in magnetization transfer prepared images is difficult due to its small size and the relatively low signal to noise ratio of these images. In this work, we applied a data-driven approach (qVBM analysis on NM-sensitive images) to ascertain PD-related reductions in NM-content in SNpc. Two clusters of significance were found in the qVBM analysis. Cluster 1 was located in the lateral-ventral tier, and cluster 2 was located medially and ventral to cluster 1. Reductions in MTC and increases in $R_2^*$ were observed in both clusters. The increase in $R_2^*$ in cluster-1 is likely associated with the disappearance of nigrosome-1 in $T_2^*$-weighted images of PD patients, given the high spatial correspondence between cluster 1 and nigrosome-1 of control subjects.

$T_2^*$- weighted images are widely used to delineate nigrosome-1 [9-11]. Consistent with earlier results [9-11], in $T_2^*$-weighted images, nigrosome-1 was seen in the majority of control subjects (27 of 28 controls) and was absent in PD (26 of 28 patients). In PD, nigrosome-1 is associated with a loss of dorsolateral nigral hyperintensity in $T_2^*$ weighted images [25]. The description of this spatial position reflects the anatomic orientation from the perspective of substantia nigra as seen in $T_2^*$-weighted images. From the anatomic orientation of substantia nigra in neuromelanin-sensitive images, nigrosome-1 is located in the lateral-ventral portion of the neuromelanin-defined SNpc (Figure 2A). Thus, loss of hyperintensity in nigrosome-1 from iron deposition and $T_2^*$-weighted hypointensity observed in the lateral-ventral SNpc [16] likely reflect identical regions of PD degeneration.

qVBM analysis on neuromelanin-sensitive images revealed two clusters of reduced neuromelanin-sensitive contrast. Cluster 1 from the qVBM analysis is located in the lateral-ventral portion of SNpc. Interestingly, this cluster shows substantial overlap in MNI-space with the nigrosome-1 control population map and reductions in MTC may indicate a loss of dopaminergic neurons in nigrosome-1 [26]. Studies examining PD-related SNpc volume or contrast loss found a reduction of neuromelanin-sensitive contrast in the posterior portion of SNpc [27, 28]. Consistent with these findings, earlier work examining lateral-ventral tier SNpc ROIs found reduced MTC in this region of PD subjects [15]. These results accord with histological studies showing the greatest amount of PD-related neuronal loss in SNpc occurs in the lateral-ventral tier of SNpc [2, 5, 6]. Further, as nigrosome-1 is in the lateral-ventral tier of SNpc, the loss of volume and MTC reported in the aforementioned imaging studies may be due to neuronal loss in nigrosome-1.

Although nigrosome-1 has received the most attention as a potential PD biomarker, nigrosomes 2-5 also experience loss of dopaminergic neurons. To date, delineation of nigrosomes 2-5 have been achieved in $T_2$-weighted images at ultra-high field strengths [10, 29]. Like nigrosome-1, nigrosomes 2-5 are seen as hyperintense regions embedded in the $T_2$-weighted substantia nigra [10, 29]. In particular, nigrosome-2 is positioned medial and ventral to nigrosome-1 in histological and imaging studies [3, 10]. Interestingly, similar spatial positioning is seen in clusters emerging from the qVBM analysis with cluster 2 positioned in a medial and ventral position to cluster 1. Thus, reductions in MTC in cluster 2 from the qVBM analysis may reflect neuronal loss in nigrosome-2. Taken together, these results suggest that reductions in neuromelanin-sensitive contrast in PD largely occur in the nigrosomes, and may manifest as reductions in SNpc volume, lateral-ventral MTC contrast, or SNpc width in PD, as reported in earlier neuromelanin-sensitive studies using explicit or incidental magnetization transfer effects [15, 27, 30-33].

There are some caveats in the present. First, UPDRS-III evaluation was conducted in the ON state and OFF state evaluations are not available. The lack of statistically significant correlations between mean $R_2^*$ and mean MTC values in qVBM clusters with clinical motor severity may be due to measurement of UPDRS-III in the ON state. Second, neuromelanin-sensitive images have low signal to noise ratio, and smoothing of the neuromelanin-sensitive image may introduce uncertainty in the spatial position of qVBM cluster 1 and reduce the overlap with nigrosome-1.

## Conclusion

In summary, we found PD-related reductions in MTC in two clusters in SNpc from the qVBM analysis and PD-related increases in $R_2^*$ were observed in both clusters. Cluster 1 spatially overlaps with nigrosome-1, as defined in $T_2^*$-weighted images. Reductions in MTC within both clusters likely reflect a loss of melanized neurons. These results suggest that reductions in MTC largely occur in the nigrosomes and volume loss shown in earlier neuromelanin-sensitive studies reflect neuromelanin depletion in nigrosomes.

**Funding Sources:**

Xiaoping Hu, Jason Langley, and Daniel Huddleston receive funding from the Michael J. Fox Foundation (MJF 10854). Daniel Huddleston is supported by NIH grant funding (NIH-NINDS 1K23NS105944-03; NIH-NIA 1R34AG056639-01A1), the American Parkinson's Disease Association Emory Center for Advanced Research, and the Lewy Body Dementia Association Emory Research Center of Excellence. Stewart Factor is supported by The Sartain Lanier Family Foundation. Recruitment of control individuals for this research was facilitated by the Emory Alzheimer's Disease Research Center (NIH-NINDS P50-AG025688). The Emory MRI facility used in this study is supported in part by funding from a Shared Instrumentation Grant (S10) grant 1S10OD016413-01 to the Emory University Center for Systems Imaging Core.

## References

[1] D.E. Huddleston, J. Langley, P. Dusek, N. He, C.C. Faraco, B. Crosson, S. Factor, X.P. Hu, Imaging parkinsonian pathology in substantia nigra with MRI, Curr Radiol Rep 6 (2018) 15.

[2] J.M. Fearnley, A.J. Lees, Ageing and Parkinson's disease: substantia nigra regional selectivity, Brain 114 ( Pt 5) (1991) 2283-301.

[3] P. Damier, E.C. Hirsch, Y. Agid, A.M. Graybiel, The substantia nigra of the human brain. I. Nigrosomes and the nigral matrix, a compartmental organization based on calbindin D(28K) immunohistochemistry, Brain 122 ( Pt 8) (1999) 1421-36.

[4] P. Damier, E.C. Hirsch, Y. Agid, A.M. Graybiel, The substantia nigra of the human brain. II. Patterns of loss of dopamine-containing neurons in Parkinson's disease, Brain 122 ( Pt 8) (1999) 1437-48.

[5] D.C. German, K. Manaye, W.K. Smith, D.J. Woodward, C.B. Saper, Midbrain dopaminergic cell loss in Parkinson's disease: computer visualization, Ann Neurol 26(4) (1989) 507-14.

[6] E. Hirsch, A.M. Graybiel, Y.A. Agid, Melanized dopaminergic neurons are differentially susceptible to degeneration in Parkinson's disease, Nature 334(6180) (1988) 345-8.

[7] D.T. Dexter, F.R. Wells, F. Agid, Y. Agid, A.J. Lees, P. Jenner, C.D. Marsden, Increased nigral iron content in postmortem parkinsonian brain, Lancet 2(8569) (1987) 1219-20.

[8] S.T. Schwarz, M. Afzal, P.S. Morgan, N. Bajaj, P.A. Gowland, D.P. Auer, The 'swallow tail' appearance of the healthy nigrosome - a new accurate test of Parkinson's disease: a case-control and retrospective cross-sectional MRI study at 3T, PLoS ONE 9(4) (2014) e93814.

[9] A.I. Blazejewska, S.T. Schwarz, A. Pitiot, M.C. Stephenson, J. Lowe, N. Bajaj, R.W. Bowtell, D.P. Auer, P.A. Gowland, Visualization of nigrosome 1 and its loss in PD: pathoanatomical correlation and in vivo 7 T MRI, Neurology 81(6) (2013) 534-40.

[10] L.A. Massey, M.A. Miranda, O. Al-Helli, H.G. Parkes, J.S. Thornton, P.W. So, M.J. White, L. Mancini, C. Strand, J. Holton, A.J. Lees, T. Revesz, T.A. Yousry, 9.4 T MR microscopy of the substantia nigra with pathological validation in controls and disease, Neuroimage Clin 13 (2017) 154-163.

[11] C. Mueller, B. Pinter, E. Reiter, M. Schocke, C. Scherfler, W. Poewe, K. Seppi, A.I. Blazejewska, S.T. Schwarz, N. Bajaj, D.P. Auer, P.A. Gowland, Visualization of nigrosome 1 and its loss in PD: pathoanatomical correlation and in vivo 7T MRI, Neurology 82(19) (2014) 1752.

[12] J. Langley, D.E. Huddleston, X. Chen, J. Sedlacik, N. Zachariah, X. Hu, A multicontrast approach for comprehensive imaging of substantia nigra, Neuroimage 112 (2015) 7-13.

[13] X. Chen, D.E. Huddleston, J. Langley, S. Ahn, C.J. Barnum, S.A. Factor, A.I. Levey, X. Hu, Simultaneous imaging of locus coeruleus and substantia nigra with a quantitative neuromelanin MRI approach, Magn Reson Imaging 32(10) (2014) 1301-6.

[14] S.T. Schwarz, T. Rittman, V. Gontu, P.S. Morgan, N. Bajaj, D.P. Auer, T1-Weighted MRI shows stage-dependent substantia nigra signal loss in Parkinson's disease, Mov Disord 26(9) (2011) 1633–38.

[15] D.E. Huddleston, J. Langley, J. Sedlacik, K. Boelmans, S.A. Factor, X.P. Hu, In vivo detection of lateral-ventral tier nigral degeneration in Parkinson's disease, Hum Brain Mapp 38(5) (2017) 2627-2634.

[16] J. Langley, D.E. Huddleston, J. Sedlacik, K. Boelmans, X.P. Hu, Parkinson's disease-related increase of T2*-weighted hypointensity in substantia nigra pars compacta, Mov Disord 32(3) (2017) 441-449.

[17] A.J. Hughes, S.E. Daniel, L. Kilford, A.J. Lees, Accuracy of clinical diagnosis of idiopathic Parkinson's disease: a clinico-pathological study of 100 cases, J Neurol Neurosurg Psychiatry 55(3) (1992) 181-4.

[18] A. Berardelli, G.K. Wenning, A. Antonini, D. Berg, B.R. Bloem, V. Bonifati, D. Brooks, D.J. Burn, C. Colosimo, A. Fanciulli, J. Ferreira, T. Gasser, F. Grandas, P. Kanovsky, V. Kostic, J. Kulisevsky, W. Oertel, W. Poewe, J.P. Reese, M. Relja, E. Ruzicka, A. Schrag, K. Seppi, P. Taba, M. Vidailhet, EFNS/MDS-ES/ENS [corrected] recommendations for the diagnosis of Parkinson's disease, Eur J Neurol 20(1) (2013) 16-34.

[19] S.M. Smith, M. Jenkinson, M.W. Woolrich, C.F. Beckmann, T.E. Behrens, H. Johansen-Berg, P.R. Bannister, M. De Luca, I. Drobnjak, D.E. Flitney, R.K. Niazy, J. Saunders, J. Vickers, Y. Zhang, N. De Stefano, J.M. Brady, P.M. Matthews, Advances in functional and structural MR image analysis and implementation as FSL, NeuroImage 23 Suppl 1 (2004) S208-19.

[20] P. Dusek, E. Bahn, T. Litwin, K. Jablonka-Salach, A. Luciuk, T. Huelnhagen, V.I. Madai, M.A. Dieringer, E. Bulska, M. Knauth, T. Niendorf, J. Sobesky, F. Paul, S.A. Schneider, A. Czlonkowska, W. Bruck, C. Wegner, J. Wuerfel, Brain iron accumulation in Wilson disease: a post mortem 7 Tesla MRI - histopathological study, Neuropathol Appl Neurobiol 43(6) (2017) 514-532.

[21] C. Langkammer, N. Krebs, W. Goessler, E. Scheurer, F. Ebner, K. Yen, F. Fazekas, S. Ropele, Quantitative MR imaging of brain iron: a postmortem validation study, Radiology 257(2) (2010) 455-62.

[22] B. Audoin, J.P. Ranjeva, M.V. Au Duong, D. Ibarrola, I. Malikova, S. Confort-Gouny, E. Soulier, P. Viout, A. Ali-Cherif, J. Pelletier, P.J. Cozzone, Voxel-based analysis of MTR images: a method to locate gray matter abnormalities in patients at the earliest stage of multiple sclerosis, J Magn Reson Imaging 20(5) (2004) 765-71.

[23] A.M. Winkler, G.R. Ridgway, M.A. Webster, S.M. Smith, T.E. Nichols, Permutation inference for the general linear model, Neuroimage 92 (2014) 381-97.

[24] J. Langley, N. He, D.E. Huddleston, S. Chen, F. Yan, B. Crosson, S. Factor, X. Hu, Reproducible detection of nigral iron deposition in 2 Parkinson's disease cohorts, Mov Disord 34(3) (2019) 416-419.

[25] P. Mahlknecht, F. Krismer, W. Poewe, K. Seppi, Meta-analysis of dorsolateral nigral hyperintensity on magnetic resonance imaging as a marker for Parkinson's disease, Mov Disord 32(4) (2017) 619-623.

[26] S. Kitao, E. Matsusue, S. Fujii, F. Miyoshi, T. Kaminou, S. Kato, H. Ito, T. Ogawa, Correlation between pathology and neuromelanin MR imaging in Parkinson's disease and dementia with Lewy bodies, Neuroradiology 55(8) (2013) 947-53.

[27] S.T. Schwarz, Y. Xing, P. Tomar, N. Bajaj, D.P. Auer, In Vivo Assessment of Brainstem Depigmentation in Parkinson Disease: Potential as a Severity Marker for Multicenter Studies, Radiology (2016) 160662.

[28] C.M. Cassidy, F.A. Zucca, R.R. Girgis, S.C. Baker, J.J. Weinstein, M.E. Sharp, C. Bellei, A. Valmadre, N. Vanegas, L.S. Kegeles, G. Brucato, U.J. Kang, D. Sulzer, L. Zecca, A. Abi-Dargham, G. Horga, Neuromelanin-sensitive MRI as a noninvasive proxy measure of dopamine function in the human brain, Proc Natl Acad Sci U S A 116(11) (2019) 5108-5117.

[29] S.T. Schwarz, O. Mougin, Y. Xing, A. Blazejewska, N. Bajaj, D.P. Auer, P. Gowland, Parkinson's disease related signal change in the nigrosomes 1-5 and the substantia nigra using T2* weighted 7T MRI, Neuroimage Clin 19 (2018) 683-689.

[30] G. Castellanos, M.A. Fernandez-Seara, O. Lorenzo-Betancor, S. Ortega-Cubero, M. Puigvert, J. Uranga, M. Vidorreta, J. Irigoyen, E. Lorenzo, A. Munoz-Barrutia, C. Ortiz-de-Solorzano, P. Pastor, M.A. Pastor, Automated Neuromelanin Imaging as a Diagnostic Biomarker for Parkinson's Disease, Mov Disord 30(7) (2015) 945-952.

[31] I.U. Isaias, P. Trujillo, P. Summers, G. Marotta, L. Mainardi, G. Pezzoli, L. Zecca, A. Costa, Neuromelanin Imaging and Dopaminergic Loss in Parkinson's Disease, Front Aging Neurosci 8 (2016) 196.

[32] K. Ogisu, K. Kudo, M. Sasaki, K. Sakushima, I. Yabe, H. Sasaki, S. Terae, M. Nakanishi, H. Shirato, 3D neuromelanin-sensitive magnetic resonance imaging with semi-automated volume measurement of the substantia nigra pars compacta for diagnosis of Parkinson's disease, Neuroradiology 55(6) (2013) 719-724.

[33] S. Reimao, P. Pita Lobo, D. Neutel, L. Correia Guedes, M. Coelho, M.M. Rosa, J. Ferreira, D. Abreu, N. Goncalves, C. Morgado, R.G. Nunes, J. Campos, J.J. Ferreira, Substantia nigra neuromelanin magnetic resonance imaging in de novo Parkinson's disease patients, Eur J Neurol 22(3) (2015) 540-6.